\def\final {1}	

\ifdefined\preprint
  \documentclass[preprint,review,12pt]{elsarticle}
\fi
\ifdefined\wordcount
  \documentclass[final,3p,times,twocolumn]{elsarticle}
\fi
\ifdefined\final
  \documentclass[final,3p,times,twocolumn]{elsarticle}
\fi

\usepackage{graphicx,stfloats}
\usepackage{color}
\usepackage[tight]{subfigure}
\usepackage[version=4]{mhchem}

\usepackage{xfrac}
\usepackage{xspace}

\graphicspath{{figs/}}
\biboptions{sort&compress}
\journal{Proceedings of the Combustion Institute}

\newcommand{\ie}{\textit{i.e.},\xspace}
\newcommand{\eg}{\textit{e.g.},\xspace}
\newcommand{\sii}{SNapS2\xspace}
\newcommand{\tit}{Spatial Dependence of Polycyclic Aromatic Compounds Growth in Counterflow Flames}

\begin{document}

\begin{frontmatter}

\title{\tit}

\author[fir]{Qi Wang}
\author[fir]{Paolo Elvati}
\author[fir]{Doohyun Kim}
\author[sec]{K. Olof Johansson}
\author[sec]{Paul E. Schrader}
\author[sec]{Hope A. Michelsen}
\author[thi]{Kevin R. Wilson}
\author[fir,for]{Angela Violi\corref{cor1}}
\ead{avioli@umich.edu}

\address[fir]{Department of Mechanical Engineering, University of Michigan, 2350 Hayward St., 2250 G.G. Brown, Ann Arbor, MI 48109-2125, United States}
\address[sec]{Combustion Research Facility, Sandia National Laboratories, Livermore, CA 94550, United States}
\address[thi]{Chemical Sciences Division, Lawrence Berkeley National Laboratory, Berkeley, CA 94720, United States}
\address[for]{Departments of Chemical Engineering, Biomedical Engineering, Macromolecular Science and Engineering, Biophysics Program, University of Michigan, Ann Arbor, MI, United States}
\cortext[cor1]{Corresponding author:}

\begin{abstract}  
The formation mechanisms of aromatic compounds in flame are strongly influenced by the chemical and thermal history that leads to their formation.
Indeed, the complex environments that characterize combustion systems do not only affect the composition of gas-phase species, but they also determine the structure and the characteristics of the soot precursors generated. 
To illustrate the importance of these effects, in this work we investigate the growth mechanisms of soot precursors in an atmospheric-pressure ethylene/oxygen/argon counterflow diffusion flame, using a combination of computational and experimental techniques. 
In diffusion flames, flow characteristics play an important role in the formation, growth, and oxidation of particles, and soot precursors are strongly affected by the flame location.
Fluid dynamics simulations and stochastic discrete modeling were employed together to identify key reaction pathways along various flow streamlines. 
The models were validated with experimental mass spectra obtained using aerosol mass spectrometry coupled with vacuum-ultraviolet photoionization. 
Results show that both the hydrogen-abstraction-acetylene-addition mechanism and oxygen-insertion reactions are responsible for the molecular growth, and their relative importance is determined by the flame conditions along the streamlines. 
Oxygenated species were detected in regions of high temperature, high atomic oxygen concentration, and relatively low acetylene abundance. 
This study also emphasizes the need to model the counterflow flame in three dimensions to capture the spatial dependence on growth mechanisms of soot precursors. 
\end{abstract}


\end{frontmatter}

\ifdefined \wordcount
\clearpage
\fi

\section{Introduction}
The evolution of gas-phase species during combustion is well known to depend on the specific physical and chemical conditions of the system. 
However, much less is known about the sensitivity of soot precursor and incipient-particle formation to small differences in thermal and chemical history.
Our group has recently reported studies indicating that the reactions responsible for soot precursor growth are highly dependent on the local conditions, for example, with oxygen chemistry overshadowing the hydrogen-abstraction-acetylene(\ce{C2H2})-addition (HACA) mechanism~\cite{1991_Frenklach_Detailedmodelingsoot,2002_Frenklach_Reactionmechanismsoot,2005_Frenklach_Migrationmechanismaromaticedge} under specific conditions~\cite{2013_Skeen_Nearthresholdphotoionizationmass,2017_Elvati_Oxygendrivensoot,2016_Johansson_Formationemissionlarge}.
HACA is the major route to molecular growth in hydrocarbon flames, largely because acetylene concentrations are generally high.
Oxygen, however, also has a great propensity to add to the edge sites of polycyclic aromatic compounds (PACs). 
Once added, the functional groups containing oxygen predominantly evolve into different types of ethers~\cite{2017_Elvati_Oxygendrivensoot,2016_Johansson_Formationemissionlarge} and, if the acetylene concentration is also high in the region where reactive oxygen adds to the PACs, furan groups are formed~\cite{2016_Johansson_Formationemissionlarge}.

Counterflow diffusion flames are ideal to study the effects of differences in local environment as the thermal and chemical properties vary substantially in space. 
This configuration also offers the opportunity to isolate the reaction zone of the diffusion flame from radical trapping, heat loss, and partial premixing at the burner~\cite{1982_Tsuji_Counterflowdiffusionflames}, making them very attractive for the analysis of flame chemistry.

In this work, we report on the main growth mechanisms of PACs as they travel in a counterflow diffusion flame. 
The competing roles of reaction models that include oxygen chemistry and HACA mechanism, are analyzed as function of temperature, chemical environment,	and flow field. 
Using a combination of computational fluid dynamics (CFD) and stochastic chemical-growth models, together with experimental data, we identify the molecular mechanisms that lead to the formation and growth of PACs along the main flow streamlines of the flame.  
 
Although counterflow flames are often studied because a one-dimensional (1D) approximation can be formulated by considering the axial centerline~\cite{1988_Smooke_comparisonnumericalcalculations,1996_Sun_structurenonsootingcounterflow,2013_Skeen_Studieslaminaropposedflow}, our results show how differences in thermal and chemical history affect the mechanisms that control molecular growth, with marked distinctions even between neighbor streamlines. High temperature, high concentration of oxygen atoms and relative low amounts of acetylene favor the formation of oxygenated-PAC.

\section{Methodology}

\subsection{Computational Fluid Dynamics}
We used three-dimensional (3D) CFD simulation, with the software CONVERGE~\cite{2015_Richards_CONVERGEv2}, to compute the temperature profile and compositions of gas-phase species present in an atmospheric-pressure ethylene/oxygen/argon counterflow diffusion flame. 
Turbulence was included using the Reynolds-averaged Navier-Stokes equations with the renormalization group k-$\epsilon$ model. 
We leveraged the counterflow flame cylindrical symmetry and used a 30$^{\circ}$ sector mesh of the burner. 
The boundary conditions, including inlet flow conditions, were taken from the experiments; the grid size inside the burner was 0.25~mm. 
The flame was initiated by volumetric energy input near the stagnation plane for a short period of time, and the simulation was performed until the flame was stabilized. 
For the gas-phase chemical model, we selected a mechanism~\cite{2000_Appel_Kineticmodelingsoot} that includes the chemistry of minor, major, and aromatic species up to pyrene. 
Results from 3D CFD simulations in terms of temperature profile and species concentrations were used as inputs to the stochastic simulations of the soot precursor growth. 

\subsection{Stochastic Modeling}
We modeled the formation of soot precursors in the counterflow flame using the recently developed \sii code [unpublished], which adopts kinetic Monte Carlo to produce a statistical representation of the soot precursor growth. 
Using generic reactions that depend on the chemical neighborhood of each atom, instead of specific species, \sii can predict the formation of soot precursors in reactive systems given an initial molecule (``seed'') and boundary conditions.
The kinetic mechanism has been steadily expanded over the years and currently includes almost 300 generic reactions (see supplemental material, \sii kinetic mechanism). 
Compared with its previous version~\cite{2014_Lai_Stochasticatomisticsimulation}, the computational performance of the \sii code has been increased by two orders of magnitude.

We ran the \sii code on both the centerline and flow streamlines (five on the fuel side and five on the oxidizer side) on the central plane of the CFD simulation.
Starting at the center of the burner, the streamlines were equally spaced and separated by 0.5~mm (see Fig.~\ref{fig:flowtemp}).
For each streamline, we simulated multiple points as the origin for \sii, running 110 simulations (100 with benzene and 10 with toluene as seeds) for each point.
The locations of the origin were identified by starting from the position where benzene (\ce{C6H6}) concentration reached \sfrac{1}{10} of its maximum value, and proceeding at intervals of 1~ms, up to 0.3~s (time origin corresponds to the outlet edge). 
Overall, we completed a total of $\approx$~218000 \sii simulations. 
In addition, mass spectra were obtained by weighting each particle lifetime and concentration of the seed at the point of origin.

\subsection{Flame setup}
The counterflow burner consists of two collinear flow tubes with an inner diameter of 12.7~mm, mounted 12~mm apart in a vertical configuration. 
The bottom tube (fuel side) delivered a mixture of 0.23~slm (standard liters per minute at 0~$^{\circ}$C and 101325~Pa) ethylene (\ce{C2H4}) and 1.10~slm argon (\ce{Ar}); the top tube (oxidizer side) supplied a mixture of 0.25~slm oxygen (\ce{O2}) and 1.20~slm \ce{Ar}. 
These flow tubes are centered between two collinear outer flow tubes that are used to deliver sheath flows of \ce{Ar} (2.30~slm on fuel side and 3.00~slm on oxidizer side).

\subsection{Gas-phase composition and Mass spectroscopy}
We determined the gas-phase composition along the centerline of this flame using flame-sampling molecular-beam mass spectrometry (VUV-MBMS). 
Detected species weighed less than $\sim$200~u, and we employed aerosol mass spectrometry (VUV-AMS) to extend the mass range of measured species to $\sim$550~u. 
We used synchrotron-generated vacuum-ultraviolet radiation for single-photon ionization. 
The experiments were performed at the Chemical Dynamics Beamline (9.0.2) at the Advanced Light Source at Lawrence Berkeley National Laboratory, Berkeley, USA. 
Details of the VUV-MBMS experimental procedures can be found in \cite{2017_Moshammer_Aromaticringformation}, in which also the basics are provided to convert the mass spectra into mole-fraction profiles as functions of the distance from the fuel outlet (DFFO).
More details are reported in the supplemental material.

The VUV-AMS measurements have been detailed in \cite{2017_Johansson_CriticalAssessmentPhotoionization}.
Briefly, a quartz probe with its tip tapered down to a 0.3-mm opening diameter was used for sampling. 
The sample was injected into the mass spectrometer through an aerodynamic lens system that focused particles into a beam that traversed three differentially pumped stages and impacted a copper target heated to 570~K. 
The sample was never exposed to atmospheric oxygen and species vaporizing from the copper target were photoionized and pulse-extracted into a linear time-of-flight drift tube at a rate of 15~kHz.

\section{Results and discussion}

\subsection{CFD results and validation}

\begin{figure*}[!b]
  \centering
  \subfigure{\label{fig:flowtemp}
    \includegraphics[width=8.2cm,trim={1cm 1.2cm 1.8cm 2.0cm},clip]{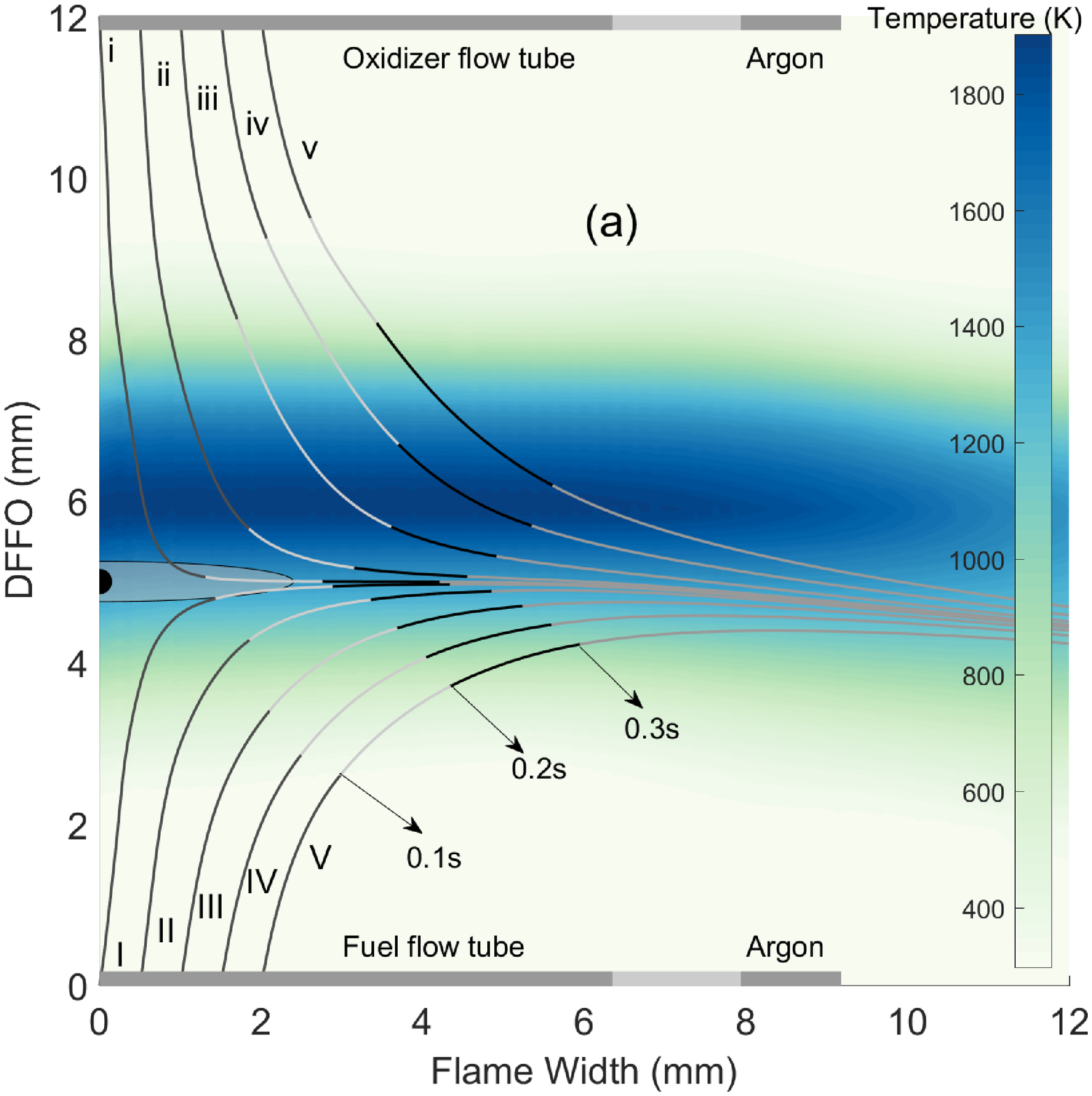} }
  \subfigure{\label{fig:traceconc}
  	\includegraphics[width=4cm,trim={0 0.8cm 0 1cm},clip]{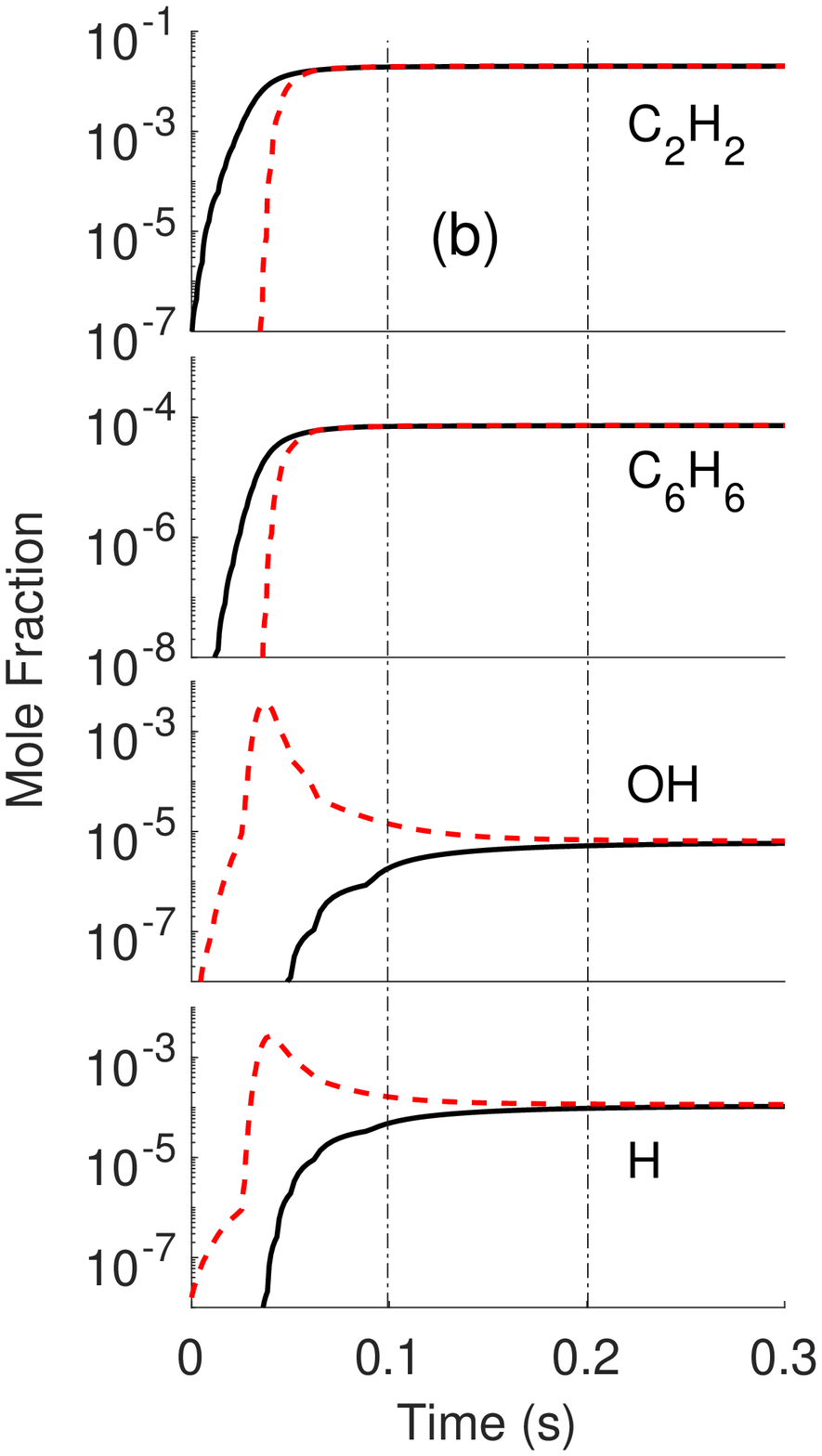} }
  \caption{\textbf{(a)} Snapshot of the central plane of the CFD simulation results. 
Background color indicates temperature and lines represent flow streamlines, labeled by Roman numerals.
At DFFO~=~5.0~mm, the black semicircle shows experimental probe size and the shadowed area is the estimated probe sampling area.
Dark gray bars at DFFO~=~0~mm and DFFO~=~12~mm display the radius of the center and outer flow tubes, separated by tube thickness in light gray.
\textbf{(b)} Time evolution of selected species concentration for \textsf{i} (black) and \textsf{I} (red dashed) streamlines.}
  \label{fig:flow.conc}
\end{figure*}

Figure~\ref{fig:flowtemp} shows the temperature contours on the central plane of the counterflow at steady state (for key species contour, see supplemental material, additional CFD data). 
The flame, located on the oxidizer side, is stabilized by the fuel diffusion across the stagnation plane, and shows a maximum temperature of 1898~K, reached at a DFFO of 6.0~mm.
The graphs in Fig.~\ref{fig:traceconc} report CFD results for the time evolutions of major gas-phase species along two streamlines, highlighting the different chemical environments between the fuel side and oxidizer side.

\begin{figure}[!htb]
	\centering
	\includegraphics[width=6.7cm,trim={0.2cm 0.8cm 2.1cm 0.6cm},clip]{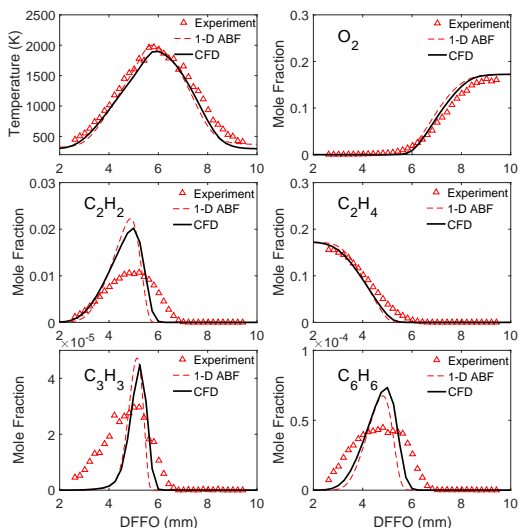}
	\caption{Temperature profile, fuel (\ce{C2H4}) and oxidizer (\ce{O2}), and other key species concentration (\ce{C2H2}, \ce{C3H3}, and \ce{C6H6}) comparison between 3D CFD (black solid line), 1D CHEMKIN (red dashed line), and experiment (triangle point) along the centerline of the counterflow flame.}
	\label{fig:species_com}
\end{figure}

Results from the 3D CFD modeling were compared with VUV-MBMS data along the centerline, as shown in Fig.~\ref{fig:species_com}. 
In the same figure, we have also included modeling results obtained employing the same kinetic mechanism as the one used for the 3D CFD runs, but in 1D approximation. The 1D results were obtained using CHEMKIN's 1D opposed-flow diffusion flames model, OPPDIF~\cite{2011_ReactionDesign_CHEMKINPRO15112}.
After shifting along the DFFO to match the peak locations, 1D CHEMKIN and 3D CFD mostly overlap for all comparisons in Fig.~\ref{fig:species_com}.
For the temperature profile, fuel (\ce{C2H4}) and oxidizer (\ce{O2}) concentrations, the CFD results predict experimental data extremely well. 
For other key species, like \ce{C2H2}, propargyl (\ce{C3H3}), and \ce{C6H6}, both the shape and the peak are well-captured by CFD simulation, but the experimental data show slightly broader curves.

\subsection{Reactivity along the centerline and streamlines}
Temperature and species profiles from CFD simulations were then used as initial and boundary conditions to the \sii code to study the evolution of soot precursors and determine their growth mechanisms as species travel through various areas of the flame. 
As initial molecules (\ie seeds) that are the starting point for the molecular growth for the \sii code, we selected benzene and toluene. 
Benzene is present in relatively high concentrations and, due to its stability, we can reasonably assume that the amount of its decomposition is negligible. 
As for toluene, we have reported in the past~\cite{2014_Lai_Stochasticatomisticsimulation} that the incorporation of odd numbered species seeds is needed to reproduce the mass peaks corresponding to odd carbon molecules.
Practical considerations preclude us from considering a larger pool of molecules, as the statistical sampling is only marginally improved by low concentration species.

\sii simulations along the centerline of the flame did not indicate any relevant growth of soot precursors.
On the fuel side, the code identified a negligible number of reactions because of the low temperature and radical concentrations. 
On the other hand, simulations along the oxidizer side displayed higher reactivity, but the chemical changes were mostly limited to the forward and reverse reactions of oxygen addition, which resulted in no significant growth.

This picture changed significantly once we performed the same analysis along the ten streamlines reported in Fig.~\ref{fig:flowtemp}. 
Molecules evolving on the oxidizer side underwent a large number of early reactions that are part of the high temperature oxidation pathways. 
These reactions are not limited to reversible addition of oxygen (as happened along the centerline), but include the formation of a large number of ethers near the oxidizer outlet that evolved along the streamlines on the oxidizer side.
On the fuel side, the molecular growth at small distances (DFFO~$<$~4.0~mm) was negligible due to the low temperature and radical concentrations. 
However, at DFFO $\approx$~5.0~mm, we detected a rapid molecular growth driven by the HACA mechanism because of the high \ce{C2H2} concentration and temperature, leading also to the formation of species containing five-membered rings formed through acetylene addition and bay-closure, as reported in previous works~\cite{2013_Kislov_FormationMechanismPolycyclic,2015_Johansson_Sootprecursorformation,2017_Johansson_Radicalradicalreactions,2015_Parker_UnexpectedChemistryReaction}.
These molecules can be detected at short distances from the fuel side outlet where a combination of thermophoresis and the low temperature are likely responsible for their condensation \cite{1997_Kang_Sootzonestructure}.

\subsection{PACs mass distribution} 
The results generated by \sii were compared with VUV-AMS.
To perform this comparison, we had to weigh the traces (molecules growth history) produced by \sii.
As a first step, we estimated the region sampled by the experimental probe, and then determined the relative concentrations of ``seed'' molecules. 
Whereas the experimental setup provides sub-mm vertical resolution along the DFFO, the horizontal width of the probe volume is uncertain.
We, therefore, assumed the probed region to be elliptical with a minor axis of 0.5~mm (slightly larger than the probe diameter 0.3~mm), and then performed a sensitivity analysis on the size of the major semi-axis of the ellipse, which was varied between 1.5~mm and 2.5~mm (see supplemental materials, probe size analysis). 
The results show that the mass spectra display the same peaks with only slight changes in the relative intensities, as we vary the length of the axis. 
As larger areas guarantee less statistical noise, we selected the final probed area (shown in Fig.~\ref{fig:flowtemp} as a shaded area) to have a major semi-axis of 2.5~mm.
For the seed molecules, the kinetic model we employed for the CFD simulations does not include toluene. 
Hence, we used the mole fractions obtained from the VUV-MBMS data to estimate the benzene/toluene seed ratio. 
The VUV-MBMS data indicate that the benzene/toluene ratio ranged mostly from 1.4 to 8.8 along the centerline. Therefore, we considered different benzene/toluene weight ratios (from 2 to 10) and analyzed their effect on the mass spectra.
The results show that changing the relative concentration of toluene only affects the relative intensity of the mass peaks corresponding to species with an odd number of carbon atoms, and we used a constant benzene/toluene ratio of 2 for all computed mass spectra. 

\begin{figure}[!htb]
\centering
\includegraphics[width=6.7cm,trim={0.5cm 0.3cm 0.9cm 0.8cm},clip]{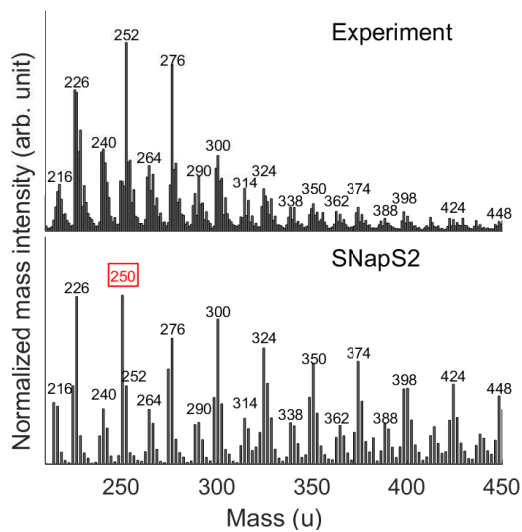}
\caption{Mass spectra at a DFFO~=~5.0~mm.
The number enclosed in a red box shows the mismatched mass peak. 
Mass intensities from \sii have a maximum relative error of 0.12\% (See supplemental material, error analysis).
Experimental mass spectra obtained with a ionization photon energy of 9.5~eV.}
\label{fig:mass5mm}
\end{figure}

Figure~\ref{fig:mass5mm} shows the simulated and the VUV-AMS data at a DFFO of 5.0~mm, which corresponds to a region on the fuel side where the VUV-AMS signal is strong and the \sii code predicts rapid precursor growth.
The comparison shows the ability of the code to reproduce mass peaks detected experimentally. These peaks stem from pure hydrocarbon species, and there is a slight preference for species containing even numbers of carbon atoms. In addition, the code reproduces the relative mass-peak intensities within mass-peak clusters very well.

VUV-AMS mass-peak intensities depend on species concentrations, transport efficiencies between the probe and the ionization region, vaporization efficiencies in the ionization region, and photoionization cross sections. 
Transport efficiencies mainly lead to reduced peak intensities on the low-mass side of the mass spectrum. 
The trends in photoionization cross sections and vaporization efficiencies are generally anti correlated, but can have significant impact over extended mass ranges. 
When all of these effects are considered, the VUV-AMS data suggest a significant decrease in species concentrations with increasing molecular mass. The VUV-AMS mass spectrum in Fig.~\ref{fig:mass5mm} has not been corrected for these effects.

The above discussion highlights a significant discrepancy between the VUV-AMS spectrum and the simulated spectrum regarding the mass-peak intensities covered in Fig.~\ref{fig:mass5mm}. This difference is likely caused by a depletion of species of heavier mass than the ones reported in Fig.~\ref{fig:mass5mm}, due to their clustering. 
Indeed, in the flame, species are subjected to different gradient-driven diffusion phenomena that are not considered by our model. 
Specifically, thermophoresis results in the diffusion of species to relatively cold parts of the flame, \eg the fuel side, where large species have a higher tendency to cluster as compared to low molecular weight compounds and are therefore effectively removed from the gas phase~\cite{1993_GOMEZ_ThermophoreticEffectsParticles}.
Other factors, like the choice of the sampling area for the computational sampling or the perturbation due to the probe, which can lower the local temperature and inhibit the molecular growth of large hydrocarbons, may also contribute to the discrepancy.

\begin{figure*}[!t]
	\centering 
	\includegraphics[width=14.4cm,trim={0 0 0 0},clip]{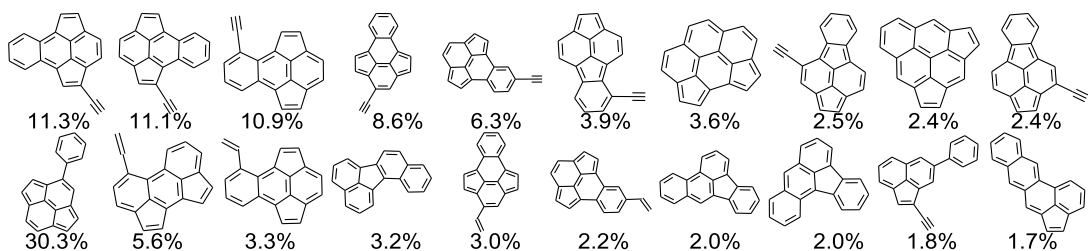}
	\caption{Top 10 most frequent species predicted by \sii for the 250~u (top row) and 252~u (bottom row) peaks shown in Fig.~\ref{fig:mass5mm}. 
The percentages indicate the relative abundance of each structure.}
	\label{fig:common_molecules}
\end{figure*}

For the cluster of signals around 250/252~u, experimental and computational data show opposite trends. 
The structures of the computed species (Fig.~\ref{fig:common_molecules}) illustrate that the peak at 250~u is composed mainly of molecules that differ only in the position of the ethynyl side chain, while the peak at 252~u is dominated by 1-phenylpyracyclene and by a relevant number of molecules with an ethenyl group.
Based on these data, a possible reason for the discrepancy could be ascribed to the preference in the simulations for the ethynyl group over the ethenyl radical. 
Another potential cause for the discrepancy is the breaking of a weak bond (\eg ether or non-covalent bond) where a 252~u fragment is formed during the heating of the sampled particles. 

\subsection{PACs reaction pathways}
\begin{figure}[!htb]
\centering
\includegraphics[width=6.7cm,trim={0.7cm 4.5cm 0.7cm 2.2cm},clip]{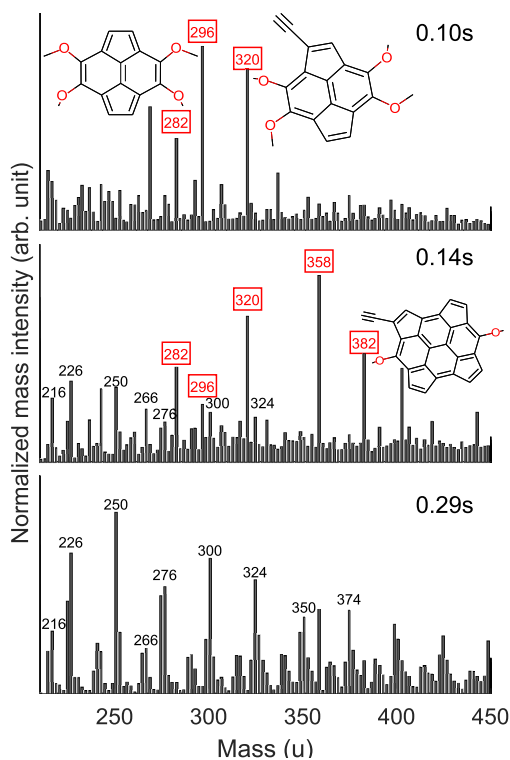}
\caption{\sii-generated mass spectra at different times along the streamline \textsf{ii}.
Boxes enclose masses of oxygenated species.
2D structures show the most probable molecules at peaks of 296~u, 320~u, and 382~u.
Mass intensities have a maximum relative error of 0.6\%, 2.8\%, and 3.5\%, from top to bottom panel, respectively.} 
\label{fig:massevolution}
\end{figure}

In our study we also identified structures containing oxygen; they are not reported because they form (on the oxidizer side) earlier in the streamline and by the time they reach the region experimentally sampled, their masses are above the interval used in Fig.~\ref{fig:mass5mm}.
To better show this effect, in Fig.~\ref{fig:massevolution} we compare the mass spectra for the \textsf{ii} streamline on the oxidizer side at three different times.

At the early stages (0.1~s), even though the molecules are just above the stagnation plane, HACA is not the dominant growth mechanism and oxygen chemistry is responsible for the majority of the growth.
These growth pathways are a consequence of the low \ce{C2H2} concentration, high temperature ($\approx$~1600~K), and elevated atomic oxygen concentration.
The resulting spectrum is not clustered in groups because of the variety of structures present in the growing molecules.
The oxygen-to-carbon (\ce{O}/\ce{C}) ratio at this stage is relatively high as the most common structures at 296~u and 320~u clearly show.
These molecular structures should be taken with a bit of caution because there is a general lack of \textit{ab initio} literature data regarding oxygen reactions rates under flame conditions~\cite{2005_Richter_DetailedmodelingPAH,2007_Chung_Insightsnanoparticleformation}.
The reliability of the reaction rates involving oxygen is thus lower than the other reactions in the \sii mechanism.  
Nevertheless, some of the small oxygenated species (\eg peaks at 168~u, 184~u, and 198~u), which are predicted by \sii in the early part of the streamlines on the oxidizer side, are also experimentally observed~\cite{2016_Johansson_Formationemissionlarge}.

At the intermediate stages (0.14~s), as \ce{C2H2} concentration increases, pathways related to HACA start to generate molecules that fill the mass region below 350~u.
The number of molecules containing oxygen and the \ce{O}/\ce{C} ratio decline because of the decreased gas-phase hydroxyl radical and atomic oxygen.
At the later stages (0.29~s), the growth is dominated by HACA, and pure hydrocarbons are mainly formed with sharp peaks at each carbon number.

\begin{figure}[bht]
\centering
\includegraphics[width=6.7cm,trim={0.5cm 0.3cm 0.9cm 0.8cm},clip]{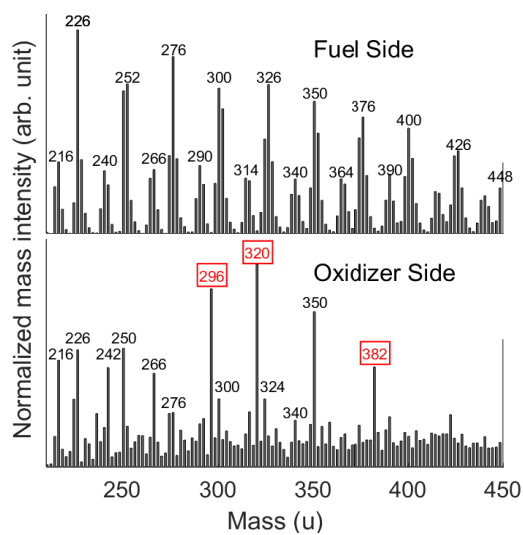}
\caption{Mass spectra of streamline \textsf{III} (fuel side) and \textsf{iii} (oxidizer side) at time = 0.29~s.}
\label{fig:massbothsides}
\end{figure}

To conclude our analysis on the effect of local environment on the growth of soot precursors, we compared the mass spectra generated along the streamline \textsf{iii} (on the oxidizer side) and streamline \textsf{III} (on the fuel side) at $\approx$ 0.3~s (Fig.~\ref{fig:massbothsides}).
In a hypothetical sampling at a DFFO of 5~mm, both these streamlines would contribute to the mass spectra, assuming that the probe sampled a small circular region of 0.5~mm in diameter.

Despite the spatial proximity, these two streamlines present rather different signals because of radically different oxygen contents in the soot precursors (\ie \ce{O}/\ce{C} ratio) and their growth pathways.
Mass spectra along streamline (\textsf{III}) on the fuel side show compounds grouped in clusters, which correspond to hydrocarbons formed by HACA model, while the other streamline (\textsf{iii}) shows a more homogeneous distribution of signals that arises from a variety of compounds, many of which contain oxygenated groups. 
Even without a detailed analysis of the differences between the compounds and reaction pathways in these two streamlines, the data clearly show how the chemistry in the counterflow flame is dependent on the exact thermal and chemical history of the streamlines that are sampled.

\section{Conclusions}
In this work, we investigated the spatial dependence on the growth pathways of soot precursors using a combination of computational and experimental tools. 
Computational fluid dynamics simulations were used to provide the temperature, minor and major concentrations of gas-phase compounds in a three-dimensional flame configuration, along various flow streamlines. 
Using stochastic simulations (\sii), we highlighted the relative importance of thermal and chemical history on the formation of polycyclic aromatic compounds. 
The \sii data, indeed, show that molecular growth is predominantly attributable to the hydrogen-abstraction-acetylene-addition mechanism, but is dominated, at early stages on the oxidizer side, by oxygen-radical reactions, since atomic oxygen concentrations are high, and \ce{C2H2} concentrations are low.
The three-dimensional fluid dynamics simulations are able to reproduce correctly the experimentally measured temperature profile and the concentrations of major species.

\sii predicts correctly the shape of the measured mass spectra and the relative intensities within peak clusters. 
Discrepancies exist in the intensities of species with high masses. A likely explanation is the lack in the model of thermophoresis.

In addition, \sii is able to shed light on the competing growth pathways as species travel along streamlines. 
While a more accurate oxygen chemistry would benefit the stochastic model, overall \sii has been proven to be an effective tool for predicting, as well as testing hypotheses on soot precursor growth in flame. 
The conclusions of this study also indicate the importance of resolving the counterflow flame as three-dimensional configuration in order to capture the spatial dependence of the molecular growth of soot precursors.

\section*{Acknowledgments}
This work was funded by the U.S. Department of Energy (DOE), Office of Basic Energy Sciences (BES), by the Single Investigator Small Group Research (SISGR), Grant No. DE-SC0002619. 
PES, HAM, and experimental expenses were funded under DOE BES, the Division of Chemical  Sciences, Geosciences, and Biosciences. 
Sandia National Laboratories is a multi-mission laboratory managed and operated by National Technology and Engineering Solutions of Sandia, LLC, a wholly owned subsidiary of Honeywell International, Inc., for the DOEs National Nuclear Security Administration under Contract No. DE-NA0003525.   
AMS measurements were performed at the ALS at LBNL. 
KRW is supported by the Department of Energy, Office of Science, Office of Basic Energy Sciences, Chemical Sciences, Geosciences, and Biosciences Division, in the Gas Phase Chemical Physics Program under Contract No. DE-AC02-05CH11231.
This research used resources of the Advanced Light Source, which is a DOE Office of Science User Facility under Contract No. DE-AC02-05CH11231.
We are grateful to Dr. Nils Hansen for letting us use his VUV-MBMS instrument and for fruitful discussions during the VUV-MBMS data evaluation. 
Dr. Kai Moshammer is also acknowledged for valuable input during the VUV-MBMS data evaluation.
Authors would also like to thank Convergent Science Inc. for providing the CFD code CONVERGE.

\section*{Supplemental material}
Supplemental material includes: \sii kinetic mechanism, additional CFD data, probe size analysis, error analysis, and gas-phase measurements.

\bibliography{Complete.bib}

\bibliographystyle{elsarticle-num-PROCI.bst}

\end{document}